\documentclass[12pt,aps,nofootinbib]{revtex4}

\usepackage{amsmath,amssymb,bm}
\usepackage[dvips]{graphicx}
\usepackage{hyperref}
\usepackage{yfonts}
\usepackage{color}

\newcommand{\beq}{\begin{eqnarray}}
\newcommand{\eeq}{\end{eqnarray}}
\renewcommand\d{\partial}

\begin{document}

\title{Spin Hall effect of gravitational waves}

\author{Naoki Yamamoto}
\affiliation{Department of Physics, Keio University, Yokohama 223-8522, Japan}

\begin{abstract}
Gravitons possess a Berry curvature due to their helicity. We derive the semiclassical equations of 
motion for gravitons taking into account the Berry curvature.
We show that this quantum correction leads to the splitting of the trajectories of right- and left-handed 
gravitational waves in curved space, and that this correction can be understood as a topological 
phenomenon. This is the spin Hall effect (SHE) of gravitational waves. 
We find that the SHE of gravitational waves is twice as large as that of light. Possible future 
observations of the SHE of gravitational waves can potentially test the quantum nature of gravitons 
beyond the classical general relativity. 
\end{abstract}
\maketitle

\section{Introduction}
One of the important predictions in Einstein's theory of general relativity is the gravitational lensing,
the deflection of light rays in a gravitational field around a massive object.
However, the classical gravitational lensing of light in curved space receives a modification, expressible
in terms of Berry curvature, due to the helicity of photons \cite{Gosselin:2006wp}.%
\footnote{The deflection of massless spinning particles in a gravitational field was previously studied in 
Refs.~\cite{Mashhoon:1975, Bailyn:1977uj, Duval:2016hxo} and that of circularly polarized light around 
rotating massive objects in Refs.~\cite{Mashhoon:1974, Frolov:2011mh} in different contexts.
Except for Ref.~\cite{Gosselin:2006wp}, however, the spin Hall effect of light in curved space in terms 
of Berry curvature has not been discussed.} In particular, such a quantum correction leads to the splitting 
of the trajectories of right- and left-handed circularly polarized light.
In the context of optics, a closely related effect of light is known in an optically inhomogeneous medium 
and is called the optical Magnus effect or spin Hall effect of light 
\cite{Liberman:1992zz, Bliokh:2004, Onoda:2004zz, Duval}.

In this paper, we show that gravitational waves exhibit the quantum spin Hall effect similarly to light, 
and that the effect is twice as large as that of light. 
Our result shows that, although the trajectories of both light and gravitational wave in the curved space 
are null geodesic and are degenerate classically in the geometric-optics limit, this degeneracy is resolved 
quantum mechanically by their helical nature. 

For this purpose, we first show that gravitons possess a Berry curvature due to their helicity, and derive 
the semiclassical equations of motion for gravitons taking into account the Berry curvature. We then 
show that the quantum correction to the gravitational lensing of gravitational waves in curved space is 
expressed by the Berry curvature, and hence, it can be understood as a topological phenomenon.
Our work demonstrates the importance of the notions of Berry curvature and topology even in the 
gravitational physics, which have been mostly investigated in the context of condensed matter physics \cite{Xiao:2010} 
and have just recently been applied in high-energy physics \cite{Son:2012wh, Stephanov:2012ki, Chen:2012ca} 
and astrophysics \cite{Yamamoto:2015gzz}.

\section{Spin Hall effect of light in gravity}
We first illustrate the spin Hall effect (SHE) of light in a gravitational field.%
\footnote{A similar result was obtained in Ref.~\cite{Gosselin:2006wp} in a different way.} 
To keep the relativistic and quantum mechanical nature apparent, we will explicitly write $\hbar$ and $c$ 
in this section.

Let us first recall the classical gravitational lensing of light in a gravitational potential $\phi({\bm x})$.
Consider the case in a weak gravitational field, where the metric is given by
\beq
{\rm d}s^2 = \left(1 + \frac{2\phi}{c^2} \right) c^2{\rm d}t^2 - \left(1 - \frac{2\phi}{c^2} \right) ({\rm d}{\bm x})^2\,.
\eeq
Here, we assume $\phi$ is static and satisfies $|\phi|/c^2 \ll 1$. Because the light propagates 
along a null geodesic, ${\rm d}s^2 =0$, the coordinate velocity of light is given by 
\beq
\label{c'}
c' \equiv \frac{|{\rm d} {\bm x}|}{{\rm d}t} = c \sqrt {\frac{1+ \frac{2\phi}{c^2}}{1-\frac{2\phi}{c^2}}}
\approx c \left(1 + \frac{2\phi}{c^2} \right)\,.
\eeq

It has been known that the null geodesic equation in the static and weak gravitational potential is 
equivalent to the geometric-optical equation of light rays in a medium with refractive index 
$n({\bm x})$ that varies depending on $\phi ({\bm x})$ \cite{Carroll}.
Equation (\ref{c'}) shows that the refractive index is given by
\beq
\label{n}
n = \sqrt{\frac{1-\frac{2\phi}{c^2}}{1+\frac{2\phi}{c^2}}} \approx 1 - \frac{2\phi}{c^2}\,.
\eeq
For $\phi<0$, $c'<c$ and $n>1$. This feature, despite being applicable to generic solutions of 
Maxwell's equations in curved space (see, e.g., Ref.~\cite{Puthoff:1999ya} and references therein), 
allows us, in particular, to describe the effects of weak gravity using purely the language of the 
geometric optics. In the following, we will consider the action of photons in the geometric-optics limit 
where the wavelength of light is much smaller than the radius of curvature of the background gravity.

We now note that the equations of motion for light in curved space are also affected by the helicity. 
In the semiclassical regime, the helical nature of photons can be expressed by the Berry connection 
or Berry curvature \cite{Liberman:1992zz, Bliokh:2004, Onoda:2004zz, Yamamoto:2017uul}. 
The action for right- and left-handed circularly polarized light in the weak gravity is given by 
\beq
\label{I}
I^{\gamma} = \int {\rm d}t \ ({\bm p} \cdot \dot {\bm x} - {\bm a}_{\bm p}^{\gamma} \cdot \dot {\bm p} - \epsilon_{\bm p})\,.
\eeq
Here, ${\bm a}_{\bm p}^{\gamma}$ is the Berry connection of photons, which is related to 
the Berry curvature ${\bm \Omega}_{\bm p}^{\gamma}$ via 
\beq
{\bm \Omega}_{\bm p}^{\gamma} \equiv {\bm \nabla}_{\bm p} \times {\bm a}_{\bm p}^\gamma
= \lambda \frac{\hat {\bm p}}{p^2}\,,
\eeq
where $\hat {\bm p} \equiv {\bm p}/|{\bm p}|$, $p \equiv |{\bm p}|$, and $\lambda$ is the helicity of photons 
($\lambda =\pm \hbar$ for right- and left-handed photons, respectively). 
As we explained above, the effect of the static gravitational potential is accounted for by the refractive 
index (\ref{n}), which modifies the energy dispersion of photons as 
\beq
\label{epsilon}
\epsilon_{\bm p} = p c' = \frac{p c}{n}\,,
\eeq
where $c'$ and $n$ are given by Eqs.~(\ref{c'}) and (\ref{n}).

The semiclassical equations of motion for the wave packet of light are obtained from 
the action (\ref{I}) as%
\footnote{For a system in a rotation ${\bm \omega}$, we have the additional Coriolis force 
$2 |{\bm p}| \dot {\bm x} \times {\bm \omega}$ in Eq.~(\ref{pdot}) in a rotation frame, which, 
combined with the Berry curvature correction in Eq.~(\ref{xdot}), leads to the 
``photonic chiral vortical effect" \cite{Yamamoto:2017uul} 
(see also Refs.~\cite{Avkhadiev:2017fxj, Zyuzin} for other derivations).}
\beq
\label{xdot}
\dot {\bm x} &=& \frac{c}{n} \hat {\bm p} + \dot {\bm p} \times {\bm \Omega}_{\bm p}^{\gamma}, \\
\label{pdot}
\dot {\bm p} &=& - \frac{2p}{c} {\bm \nabla} {\phi}\,.
\eeq
In the context of the geometric optics, the second term in Eq.~(\ref{xdot}) is called the 
optical Magnus effect \cite{Liberman:1992zz, Bliokh:2004}. On the other hand, 
Eq.~(\ref{pdot}) represents the (classical) gravitational lensing effect in the gravitational 
potential $\phi$. Inserting Eq.~(\ref{pdot}) into Eq.~(\ref{xdot}), we have
\beq
\label{SHE1}
\dot {\bm x} = \frac{c}{n} \hat {\bm p} - \frac{2 \lambda}{c} \bm \nabla \phi \times \frac{\hat {\bm p}}{p}\,.
\eeq
The second term represents the quantum spin Hall effect of light induced by the background curved 
geometry. The trajectory of light is shifted in the direction perpendicular to both ${\bm \nabla} \phi$ 
and the classical trajectory $\hat {\bm p}$, and in particular, the trajectories of right- and left-handed 
circularly polarized light are separated. This effect originates from the interplay between general 
relativity and the helical nature of right- or left-handed photons. 

As an example, consider the Newtonian potential at a distance $r$ from a point mass $M$:
\beq
\label{phi_Newton}
\phi(r) = -\frac{GM}{r}\,,
\eeq
where $G$ is the universal gravitational constant. In this case, Eq.~(\ref{SHE1}) reduces to
\beq
\label{SHE2}
\dot {\bm x} = \frac{c}{n} \hat {\bm p} - \frac{2 \lambda GM}{c} \frac{\hat {\bm r}}{r^2} \times \frac{\hat {\bm p}}{p}\,.
\eeq
Equation (\ref{SHE2}) shows that the SHE becomes larger as $M$ increases and 
as $p$ decreases (for fixed $r$). Thus, the SHE becomes particularly relevant for 
electromagnetic waves with long wavelength around massive astrophysical objects. 

It is straightforward to get the generic kinetic theory for photons in the weak gravitational field.
By inserting Eqs.~(\ref{pdot}) and (\ref{SHE1}) into the kinetic equation,
\beq
\label{kinetic1}
\frac{\d f_{\lambda}}{\d t} + \dot {\bm x} \cdot \frac{\d f_{\lambda}}{\d {\bm x}} + \dot {\bm p} \cdot \frac{\d f_{\lambda}}{\d {\bm p}} = C[f_{\lambda}]\,,
\eeq
where $f_{\lambda} = f_{\lambda}(t, {\bm x}, {\bm p})$ is the distribution function of photons 
with helicity $\lambda$ and $C[f_{\lambda}]$ is the collision term, we get
\beq
\label{kinetic2}
\frac{\d f_{\lambda}}{\d t} 
+ \left(\frac{c}{n} \hat {\bm p} - \frac{2 \lambda}{c} \bm \nabla \phi \times \frac{\hat {\bm p}}{p} \right) \cdot \frac{\d f_{\lambda}}{\d {\bm x}}
- \frac{2p}{c} {\bm \nabla} {\phi} \cdot \frac{\d f_{\lambda}}{\d {\bm p}} = C[f_{\lambda}]\,.
\eeq
This equation describes the time evolution of right- and left-handed photons for any given 
(weak and static) gravitational field $\phi$.

\section{Spin Hall effect of gravitational waves}
In this section, we consider the SHE of gravitational waves in the semiclassical regime.
At the classical level, the gravitational wave travels along a null geodesic, and hence, 
the propagation of the gravitational wave in curved space is described by the same 
geometric-optical equation as the case of light \cite{MTW}. We will show that gravitons 
possess a Berry curvature in the semiclassical regime, leading to the SHE of gravitons 
in curved space, similarly to that of photons above.

\subsection{Generalized Weyl equation with any helicity}
Let us briefly review the generalized Weyl equation for massless fields with any spin 
\cite{Bacry:1975di, Gersten:2011zz}, and then we apply it to spin-2 gravitons. 
In the following, we use the natural units $\hbar = c = 1$ for simplicity, 
unless stated otherwise.

We first recall the representation of the Poincar\'e algebra for massless fields. 
The Poincar\'e symmetry consists of the space-time translations generated by the 
energy-momentum vector $p^{\mu}$ and the Lorentz transformations generated by 
$M^{\mu \nu}$. In 3+1 space-time dimensions, there are two Casimir operators that
commute with $p^{\mu}$ and $M^{\mu \nu}$: 
$p^2 = p^{\mu} p_{\mu}$ and $W^2 = W^{\mu} W_{\mu}$, where $W^{\mu}$ is the 
Pauli-Lubanski vector defined by
\beq
\label{W}
W^{\mu} = -\frac{1}{2}\epsilon^{\mu \nu \alpha \beta} p_{\nu} M_{\alpha \beta}\,.
\eeq
Because the contribution of the orbital angular momentum vanishes due to the 
antisymmetry with $p_{\nu}$, Eq.~(\ref{W}) can also be rewritten as
\beq
\label{W2}
W^{\mu} = -\frac{1}{2}\epsilon^{\mu \nu \alpha \beta} p_{\nu} S_{\alpha \beta}
= -p_{\nu} \tilde S^{\mu \nu}\,, \qquad 
\tilde S^{\mu \nu} \equiv \frac{1}{2}\epsilon^{\mu \nu \alpha \beta} S_{\alpha \beta}\,,
\eeq
where $S^{\mu \nu}$ is the spin tensor, whose components are $S^{ij} = \epsilon^{ijk} S^k$ and 
$S^{0i}=i S^i$ with $S^i$ being the spin vector.

Let us now introduce the ($2|\lambda|+1$)-component massless field $\psi$ with helicity 
$\lambda$, which satisfies the equation
\beq
W^{\mu} \psi = \lambda p^{\mu} \psi,
\eeq
according to Wigner's result \cite{Wigner:1939cj}. Using Eq.~(\ref{W2}), this equation can be written as
\beq
(\tilde S^{\mu \nu}p_{\nu} + \lambda p^{\mu})\psi = 0.
\eeq
The temporal ($\mu=0$) and spatial ($\mu = i$) components of this equation are
given by 
\begin{gather}
\label{Weyl}
({\bm S} \cdot {\bm p} - \lambda p^0) \psi = 0,  \\
\label{sub1}
({\bm S} p^0 + i {\bm S} \times {\bm p} - \lambda {\bm p}) \psi =0,
\end{gather}
respectively. 
Equation (\ref{Weyl}) is the generalized Weyl equation for massless field with helicity 
$\lambda$, and Eq.~(\ref{sub1}) is the subsidiary condition \cite{Bacry:1975di, Gersten:2011zz}. 
The generalized Weyl Hamiltonian corresponding to the wave equation~(\ref{Weyl}) is
\beq
\label{H}
H = \frac{1}{\lambda}{\bm S} \cdot {\bm p}\,.
\eeq

Note that three components of Eq.~(\ref{sub1}) are not independent. 
To see this, we first eliminate $p^0$ in Eq.~(\ref{sub1}) using Eq.~(\ref{Weyl}) to get
\beq
\label{sub2}
\left[\frac{1}{\lambda} {\bm S} ({\bm S} \cdot {\bm p}) + i {\bm S} \times {\bm p} - \lambda {\bm p} \right] \psi =0\,.
\eeq
It is then easy to check that the inner product of the left-hand side of Eq.~(\ref{sub2}) with ${\bm S}$ 
vanishes, meaning that only one of three components in Eq.~(\ref{sub2}) is independent.
Without loss of generality, we take the $z$-component of Eq.~(\ref{sub2}), 
\beq
\label{sub_z}
\left[\frac{1}{\lambda} S_z ({\bm S} \cdot {\bm p}) + i \left(S_x p_y - S_y p_x \right) - \lambda p_z \right] \psi =0\,,
\eeq
as the subsidiary condition to Eq.~(\ref{Weyl}).

In particular, for gravitons with $\lambda = 2$, a matrix representation of ${\bm S}$ is
\begin{equation}
S_x = 
\begin{pmatrix}
0 & 1 & 0 & 0 & 0 \\
1 & 0 & \sqrt{\frac{3}{2}} & 0 & 0 \\
0 & \sqrt{\frac{3}{2}} & 0 & \sqrt{\frac{3}{2}} & 0 \\
0 & 0 & \sqrt{\frac{3}{2}} & 0 & 1 \\
0 & 0 & 0 & 1 & 0 \\
\end{pmatrix}\,, \quad 
S_y = i
\begin{pmatrix}
0 & -1 & 0 & 0 & 0 \\
1 & 0 & -\sqrt{\frac{3}{2}} & 0 & 0 \\
0 & \sqrt{\frac{3}{2}} & 0 & -\sqrt{\frac{3}{2}} & 0 \\
0 & 0 & \sqrt{\frac{3}{2}} & 0 & -1 \\
0 & 0 & 0 & 1 & 0 \\
\end{pmatrix}\,, \quad 
\end{equation}
and $S_z = {\rm diag}(2,1,0,-1,-2)$.

\subsection{Path integral formulation of gravitons in flat space}
\label{sec:path}
Once the wave equation (\ref{Weyl}) for gravitons is obtained, the semiclassical equations of 
motion for gravitons taking into account the Berry curvature can be derived in a way analogous 
to Ref.~\cite{Stephanov:2012ki} for chiral fermions. The important difference in our case is, 
however, the additional constraint (\ref{sub_z}), which selects out the physical degrees of 
freedom of gravitons with helicity $\pm 2$, similarly to the situation of photons with helicity 
$\pm 1$ \cite{Yamamoto:2017uul}.

Let us consider the path integral quantization for the Hamiltonian (\ref{H}) for $\lambda=2$:
\beq
\label{path}
Z = \int {\cal D}x {\cal D}p {\cal P} e^{iI}\,, \qquad
I = \int{\rm d}t\,({\bm p} \cdot \dot {\bm x} - H) \,,
\eeq
where ${\cal P}$ denotes the path-ordered product of the matrices $\exp(-i H \Delta t)$ over the 
path in the phase space. The eigenvalues of $H$ are given by 
$\pm |\bm p|$, $\pm \frac{1}{2}|\bm p|$, and $0$, and $H$ can be diagonalized using a unitary 
matrix $V_{\bm p}$ as
\begin{equation}
V_{\bm p}^{\dag} H V_{\bm p} = |\bm p| \Gamma,
\qquad \Gamma \equiv \frac{1}{2}S_z\,. 
\end{equation} 
The eigenstates of the eigenvalues $-|{\bm p}|$ and $-\frac{1}{2}|{\bm p}|$ have the negative 
energies and are not physical. Also, one can check that the eigenstates of the eigenvalues 
$0$ and $\frac{1}{2}|{\bm p}|$ are forbidden by the subsidiary condition (\ref{sub_z}).
Therefore, we have only one physical eigenstate with the eigenvalue $|{\bm p}|$, which 
corresponds to helicity $\lambda=2$.

Following the procedure in Refs.~\cite{Stephanov:2012ki, Yamamoto:2017uul}, we diagonalize 
the matrix in the exponential factor of the path integral (\ref{path}) at each point of the trajectory as
\begin{align}
\label{factor}
\cdots \exp \Bigl( - \frac{i}{2}{\bm S} \cdot {\bm p}_2 \Delta t \Bigr)
\exp \Bigl( - \frac{i}{2} {\bm S} \cdot {\bm p}_1 \Delta t \Bigr) \cdots 
& = \cdots V_{{\bm p}_2} \exp(-i |{\bm p}_2| \Gamma \Delta t) V_{{\bm p}_2}^{\dag} 
V_{{\bm p}_1} \exp(-i |{\bm p}_1| \Gamma \Delta t) V_{{\bm p}_1}^{\dag} \cdots 
\nonumber \\
& = \cdots V_{{\bm p}_2} \exp(-i |{\bm p}_2| \Gamma \Delta t) \exp(-i \hat {\bm a}^{\rm G}_{\bm p} \cdot \dot {\bm p} \Delta t)
\nonumber \\
& \qquad \times \exp(-i |{\bm p}_1| \Gamma \Delta t) V_{{\bm p}_1}^{\dag} \cdots \,,
\end{align}
where $\hat {\bm a}^{\rm G}_{\bm p} \equiv i V_{\bm p}^{\dag} {\bm \nabla}_{\bm p} V_{\bm p}$. 
In deriving the last equation above, we used 
\beq
\label{V}
V_{{\bm p}_2}^{\dag} V_{{\bm p}_1} \approx \exp(-i \hat {\bm a}^{\rm G}_{\bm p} \cdot \Delta {\bm p})
= \exp(-i \hat {\bm a}^{\rm G}_{\bm p} \cdot \dot {\bm p} \Delta t)
\eeq
for sufficiently small $\Delta {\bm p} \equiv {\bm p}_2 - {\bm p}_1$.

Taking the semiclassical limit where off-diagonal components of $\hat {\bm a}^{\rm G}_{\bm p}$ are 
ignored,%
\footnote{This approximation is justified when $|\dot {\bm p}| \ll |{\bm p}|^2$ \cite{Stephanov:2012ki, Yamamoto:2017uul}. 
This condition is indeed satisfied for the semiclassical regime of gravitons in the weak background gravitational field $\phi$.}
we obtain the semiclassical action for gravitons in the flat space:
\beq
\label{I2}
I^{\rm G} = \int {\rm d}t\,({\bm p} \cdot \dot {\bm x} - {\bm a}^{\rm G}_{\bm p} \cdot \dot {\bm p} - \epsilon_{\bm p}),
\eeq 
where $\epsilon_{\bm p} = |{\bm p}|$ is the energy dispersion and 
${\bm a}^{\rm G}_{\bm p} \equiv [{\hat {\bm a}}_{\bm p}]_{11}$ is the Berry connection in momentum space
that originates from the helicity of gravitons. From the definition of ${\bm a}^{\rm G}_{\bm p}$ above, 
one finds the Berry curvature of gravitons as 
\beq
\label{curvature}
{\bm \Omega}^{\rm G}_{\bm p} \equiv {\bm \nabla}_{\bm p} \times {\bm a}^{\rm G}_{\bm p}
= \lambda \frac{\hat {\bm p}}{|\bm p|^2}\,,
\eeq
where $\lambda$ is the helicity of gravitons. 
This corresponds to the fictitious magnetic field of the magnetic monopole with charge,
\beq
\label{k}
k = \frac{1}{4\pi} \int {\bm \Omega}^{\rm G}_{\bm p} \cdot {\rm d}{\bm S} = \lambda \,.
\eeq
Note that Eq.~(\ref{k}) is a general relation connecting the helicity $\lambda$ to the topological charge $k$, 
which is applicable not only to gravitons, but also to photons and chiral fermions: 
$k = \pm 2$ for gravitons with $\lambda = \pm 2$ (as shown here), 
$k = \pm 1$ for photons with $\lambda = \pm 1$ \cite{Yamamoto:2017uul}, and 
$k = \pm \frac{1}{2}$ for chiral fermions with $\lambda = \pm \frac{1}{2}$ \cite{Son:2012wh, Stephanov:2012ki, Chen:2012ca}. 
This extended universal relation is one of our main results.

\subsection{Semiclassical equations of motion for gravitons in curved space}
\label{sec:eom}
In the weak and static gravitational potential $\phi({\bm x})$, the energy dispersion of gravitons
is modified as Eq.~(\ref{epsilon}) in the same way as photons, where $n$ is the 
``refractive index" of space in Eq.~(\ref{n}). In this case, the action of the graviton is 
given by Eq.~(\ref{I2}) with $\epsilon_{\bm p}$ being replaced by Eq.~(\ref{epsilon}). 
Then, the semiclassical equations of motion for gravitons become
\beq
\label{xdot2}
\dot {\bm x} &=& \frac{c}{n} \hat {\bm p} + \dot {\bm p} \times {\bm \Omega}^{\rm G}_{\bm p}, \\
\label{pdot2}
\dot {\bm p} &=& - \frac{2p}{c} {\bm \nabla} {\phi}\,.
\eeq
The second term on the right-hand side of Eq.~(\ref{xdot2}) is the ``Lorentz force" in 
momentum space, which may be regarded as the gravitational Magnus effect.

From the two equations above, we obtain Eq.~(\ref{SHE1}) with helicity $\lambda = \pm 2$. 
The second term of this equation is the SHE of the gravitational wave, which is twice as large 
as that of light because of the difference of helicity. This means that, although the trajectory of the 
gravitational wave in curved space is classically the same as that of light in the geometric-optics 
limit, this degeneracy of trajectories is lifted by the quantum effects.
Although this quantum correction looks qualitatively similar to the spin-curvature coupling 
appearing in the Mathisson-Papapetrou-Dixon equations \cite{Mathisson, Papapetrou, Dixon}
(see also Refs.~\cite{Mashhoon:1975, Bailyn:1977uj, Duval:2016hxo}), 
our result clarifies its topological nature for the first time, to the best of our knowledge. 
In particular, it reveals the universality of topological phenomena between gravitons, photons, 
and chiral fermions \cite{Son:2012wh, Stephanov:2012ki, Chen:2012ca} through the relation (\ref{k}), 
in background gravitational or electromagnetic fields.

Similarly to the case of photons, the kinetic equation for gravitons in the weak gravity is given 
by Eq.~(\ref{kinetic2}) with $f_{\lambda} = f_{\lambda}(t, {\bm x}, {\bm p})$ being replaced by 
the distribution function of gravitons.

\section{Discussions}
\label{sec:discussion}
In this paper, we derived the quantum correction to the gravitational lensing of gravitational 
waves in curved space. In particular, this correction causes the splitting of the trajectories of 
right- and left-handed circularly polarized gravitational waves. 

To quantify the SHE in gravity, consider an electromagnetic wave or a gravitational wave 
passing at a distance $r$ ($\gg R_{\rm s}$) from a Schwarzschild black hole with mass $M$ 
as an example, where $R_{\rm s} = 2GM/c^2$ is the Schwarzschild radius.%
\footnote{For a study of the {\it classical} gravitational lensing by a Schwarzschild black hole,
see Ref.~\cite{Virbhadra:1999nm}.}
From Eq.~(\ref{SHE2}), the relative magnitude of the local shift due to the SHE, compared 
with the classical trajectory, is written as 
\beq
A = \frac{n|\lambda|}{2\pi \hbar} \left(\frac{R_{\rm s}}{r}\right)^{\! 2} \! \left(\frac{\ell}{R_{\rm s}} \right)\,, 
\eeq
where $\ell$ is the wavelength. 
This relation shows that the SHE becomes more relevant as $\ell$ becomes larger, as long as 
the semiclassical approximation is valid. For example, for a black hole with solar mass $M=M_{\odot}$, 
the relative magnitude is $A \sim 10^{-3}$ for $r \sim 5 R_{\rm s}$ and $\ell \sim 300 \ {\rm m}$. 

It does not seem feasible to observe the SHE both for electromagnetic and gravitational waves by 
the current detectors. However, possible future observations of the SHE of gravitational waves, in particular, 
could test the quantum nature of gravitons beyond the classical general relativity.

\section*{ACKNOWLEDGEMENTS}
The author thanks M.~Hongo and X.~G.~Huang for useful conversations. 
This work was supported by JSPS KAKENHI Grant No.~16K17703 and MEXT-Supported 
Program for the Strategic Research Foundation at Private Universities, ``Topological Science" 
(Grant No.~S1511006).

\end{document}